# Bacterial pathways for the biosynthesis of ubiquinone

Sophie Saphia Abby[1], Katayoun Kazemzadeh[1$], Charles Vragniau[1$], Ludovic Pelosi[1]*, Fabien Pierrel[1]*

[1] Univ. Grenoble Alpes, CNRS, CHU Grenoble Alpes, Grenoble INP, TIMC-IMAG, F-38000 Grenoble, France

$ authors contributed equally to the work

* Correspondence: Fabien Pierrel and Ludovic Pelosi

TIMC Laboratory, UMR5525, Jean Roget building, Domaine de la Merci, 38700 La Tronche, France
Tél +33 4 76 63 74 79

fabien.pierrel@univ-grenoble-alpes.fr, ludovic.pelosi@univ-grenoble-alpes.fr

**Highlights**

- Ubiquinone is a crucial component of bacterial bioenergetics
- The bacterial pathways to produce ubiquinone are highly diverse
- Ubiquinone is produced in anoxic conditions by a dioxygen-independent pathway
- Ubi-, mena- and plasto-quinone biosynthetic pathways are evolutionary related

**Abstract**

Ubiquinone is an important component of the electron transfer chains in proteobacteria and eukaryotes. The biosynthesis of ubiquinone requires multiple steps, most of which are common to bacteria and eukaryotes. Whereas the enzymes of the mitochondrial pathway that produces ubiquinone are highly similar across eukaryotes, recent results point to a rather high diversity of pathways in bacteria. This review focuses on ubiquinone in bacteria, highlighting newly discovered functions and detailing the proteins that are known to participate to its biosynthetic pathways. Novel results showing that ubiquinone can be produced by a pathway independent of dioxygen suggest that ubiquinone may participate to anaerobiosis, in addition to its well established role for aerobiosis. We also discuss the supramolecular organization of ubiquinone biosynthesis proteins and we summarize the current understanding of the evolution of the ubiquinone pathways relative to those of other isoprenoid quinones like menaquinone and plastoquinone.

## 1) Introduction

Isoprenoid quinones are central to bioenergetics as they constitute obligate electrons and protons shuttles in the respiratory chains of most organisms. These molecules are composed of a quinone head group (in most cases a naphto- or a benzo- ring) to which is attached a polyprenyl tail with a length that varies depending on the organisms (from 4 to 14 isoprene units, indicated as subscript, $Q_{4-14}$) [1]. A specific number of isoprene units characterizes the quinone pool ($Q_n$) of a given species, but several lower abundance isoprenologs, typically $Q_{n-1}$ and $Q_{n+1}$, are also usually produced [2,3]. The polyprenyl tail is hydrophobic and localizes isoprenoid quinones inside membranes. The polar head group is the functional part of the molecule and undergoes a two-steps redox chemistry between quinone (oxidized) and quinol (reduced) forms [1]. Historically, isoprenoid quinones have been used as chemotaxonomic markers [2,3] and more recently, quinone profiles served as markers of bacterial communities in complex ecosystems [4,5].

Isoprenoid quinones are classified based on the nature of the head group and also according to their midpoint redox potential. Menaquinone (MK) belongs to naphtoquinones, whereas ubiquinone (UQ), plastoquinone (PQ) and rhodoquinone (RQ) are benzoquinones. Classically, MK and RQ are considered low potential quinones (E' ~ -70 mV), whereas UQ and PQ are high potential quinones (E' ~ +100 mV) [6]. The redox potential of quinones determines the protein partners with which they functionally interact in respiratory chains. Excellent reviews have covered the taxonomic distribution, the functions and the biosynthesis of isoprenoid quinones [1,7] and the case of RQ is discussed in details by Shepherd and colleagues in this issue of BBA-Bioenergetics [8].

MK, which is present in most bacterial phyla and in archaea, was proposed to have been a component of the bioenergetic toolbox of the last universal common ancestor (LUCA) [6,9]. In contrast, UQ evolved later and is restricted to specific classes of proteobacteria ($\alpha$, $\beta$, $\gamma$) and to eukaryotes, in which UQ participates to oxidative phosphorylation in mitochondria. Much attention has been paid to the biosynthesis and functions of UQ (also called coenzyme Q) in eukaryotes, and these topics - including the pathologies resulting from coenzyme Q deficiencies in humans - have been covered recently in authoritative reviews [7,10–13].

In contrast, an update is needed for UQ in bacteria. The historical model to study bacterial UQ biosynthesis has been *Escherichia coli* and several recent discoveries advanced our understanding of the UQ pathway in this species. At the same time, studies on other bacterial models revealed substantial differences with the *E. coli* pathway and highlighted an unsuspected diversity of solutions evolved by bacteria to synthesize UQ. The numerous sequences of bacterial genomes now available in public databases also appear to be a very relevant source of information with this respect. In this review, we summarize the recent results obtained on bacterial UQ biosynthesis and functions, and we emphasize how they advanced our current understanding of the field.

**2) New functions for UQ in bacteria**

The functions of UQ related to respiration, gene regulation and oxidative stress have been reviewed elsewhere [14,15] and will not be covered here. In 2014, Sevin and Sauer reported that UQ promotes tolerance to osmotic stress in *E. coli* [16]. The authors showed that the growth of a UQ-deficient strain was impaired when the medium contained high concentrations of salt. Furthermore, they observed a ~100 fold increase of the UQ content in response to osmotic stress [16]. In such conditions, UQ represented ~1% of the lipids constituting the plasma membrane of *E. coli*. UQ and structural analogs had a stabilizing effect on liposomes, which led the authors to propose that the polyprenyl tail of UQ mediates a mechanical stabilization of the plasma membrane that likely explains the osmoprotective effect observed *in vivo* [16]. However, these results have been challenged recently [17]. Indeed, a new study suggests that the impaired growth of UQ-deficient *E. coli* cells at high osmotic pressure was simply caused by the compromised function of the respiratory chain, which affected the proton-solute symporter ProP [17]. ProP mediates the uptake of zwitterionic osmolytes such as proline and glycine betaine, and requires high proton-motive force for function. As the proton gradient is compromised in the absence of UQ, the function of ProP is impaired, impacting osmotic regulation [17]. The authors also found that UQ amounted to ~1% of the total lipids, but the UQ content was not significantly modulated by the osmotic pressure of the growth medium [17], consistent with unpublished results from several laboratories (personal communications of David Pagliarini and Gilles Basset) and ours. Overall, it appears that UQ levels do not respond to osmotic stress and that the decreased tolerance to osmotic stress observed in UQ-deficient *E. coli* cells results from an indirect effect of the inactivation of the respiratory chain [17]. Even though multiple *in vitro* studies reported that UQ modifies the mechanical and physical properties of liposomes ([16,18] and references therein), sometimes at UQ contents hardly compatible with biological levels, the direct impact of UQ on the properties of membranes does not seem relevant for protection against osmotic stress *in vivo*.

A new contribution of UQ to cell metabolism was described by Chaba and colleagues who showed that UQ is required by *E. coli* to grow on medium containing long-chain fatty acids (LCFAs) as a carbon source [19]. Interestingly, mutants with intermediate UQ levels (15-20% compared to wild-type) grew normally on various non-fermentable carbon sources but not on the LCFA oleate. Thus, in addition to its role as an electron shuttle in the respiratory chain, UQ has another function in oleate metabolism. The authors suggested that the antioxidant function of the reduced form of UQ was important based on the observations that the level of reactive oxygen species (ROS) increased in cells metabolizing oleate, and that supplementation with antioxidants improved growth and decreased ROS levels of UQ-deficient mutants in oleate-containing medium [19]. Remarkably, UQ levels increased ~1.8 fold in cells metabolizing oleate. Overall, the authors proposed that UQ is the preponderant antioxidant system during LCFA degradation and acts to mitigate ROS production by the acyl-CoA dehydrogenase FadE. In this regard, UQ might be particularly important for pathogenic bacteria that use LCFAs derived from host tissues as their main nutrient [19].

### 3) UQ biosynthesis in *E. coli*

#### 3.1 Biochemical steps of the classical pathway and enzymes involved

Over a period of several decades, the UQ biosynthetic pathway has been extensively studied in *E. coli*, a bacterium that synthesizes $UQ_8$ as its main isoprenolog. Nowadays, the pathway is known to require twelve proteins (UbiA to UbiK and UbiX), most of them being involved in reactions that modify the aromatic ring originating from 4-hydroxybenzoic acid (4-HB) (Figure 1). UbiC is the first committed enzyme in the biosynthesis of $UQ_8$, catalyzing the removal of pyruvate from chorismate to produce 4-HB [20]. Then, the membrane-bound UbiA prenylates 4-HB using octaprenyl diphosphate as a side chain precursor [21]. Recent three dimensional (3D) structures of two members in the UbiA superfamily [22,23] revealed an all α-helical structure that is completely different from the α/β barrel structure of soluble aromatic prenyltransferases, in agreement with a catalysis that occurs in lipid bilayers. Both UbiA homologs contain nine transmembrane helices arranged counterclockwise in a U-shape presenting a large central cavity with an opening assimilated to a lateral portal that is largely buried in the membrane. It was proposed that this lateral portal may facilitate the binding of long-chain isoprenyl diphosphate substrates, the prenylated products being directly released into membranes through this portal [22,23]. We note that the two crystallized UbiA homologs belong to archaeal species and as such do not participate to UQ biosynthesis. However, given the conservation of important catalytic residues with *E. coli* UbiA [23], we believe that the structural insights provided by these structures are largely applicable to UbiA family members involved in UQ biosynthesis.

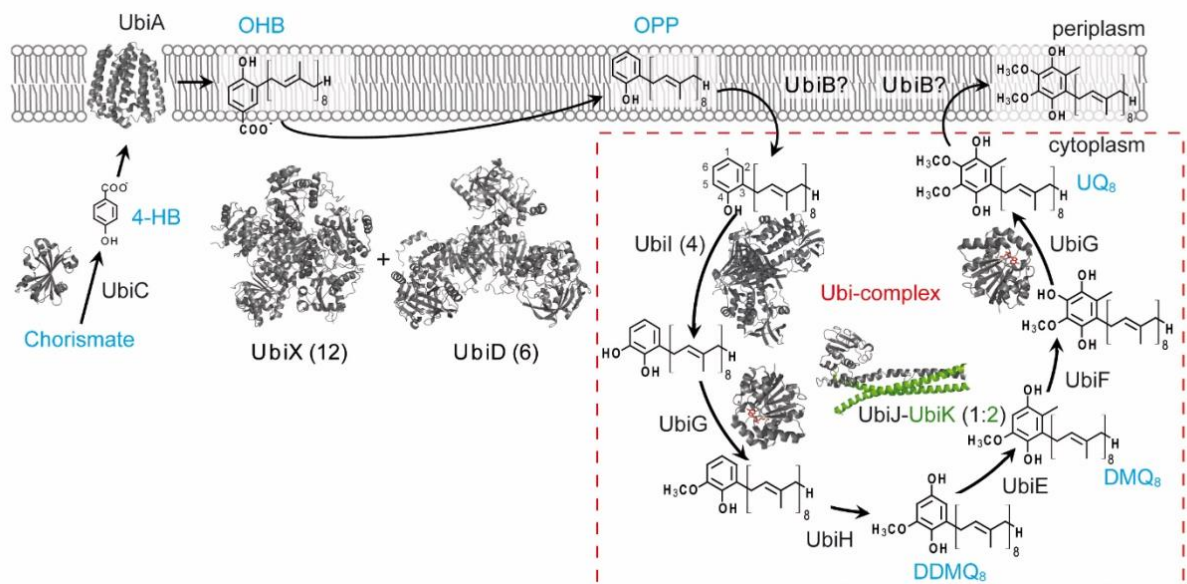

**Figure 1**: Model of UQ$_8$ biosynthesis in *E. coli* and supramolecular organization of the pathway. The names of precursors and intermediates are indicated in blue. The red dotted rectangle delimits the Ubi-complex, which is composed of UbiE to UbiK proteins and encompasses the last six reactions of the pathway. The numbering of the aromatic carbon atoms is shown on OPP. Abbreviations used are: 4-HB, 4-hydroxybenzoate; OHB, octaprenyl-4-hydroxybenzoate; OPP, octaprenyl phenol; DDMQ$_8$, C2-demethyl-C6-demethoxy-ubiquinone 8; DMQ$_8$, C6-demethoxy-ubiquinone 8; UQ$_8$, ubiquinone 8. The 3D-structures of UbiC (PDB ID: 1G81), UbiI (PDB ID: 4K22), UbiD (PDB ID: 5M1B) and UbiG (PDB ID: 4KDR) correspond to the proteins from *E. coli* and the 3D-structures of UbiX (PDB ID: 4RHE) and UbiA (PDB ID: 4TQ5) correspond to the proteins from *Pseudomonas aeruginosa* and *Archaeoglobus fulgidus*, respectively. The UbiJ monomer is colored grey and the UbiK dimer is colored green in the model of the *E. coli* UbiK–UbiJ 2:1 heterotrimer complex [24]. The oligomerization state of the 3D-models is indicated in brackets when it is greater than one. The ribbon diagrams were drawn using PyMOL (DeLano Scientific LLC).

Following its prenylation, 4-HB is decarboxylated by the UbiD-UbiX system, which consists of the decarboxylase UbiD and its associated flavin prenyltransferase UbiX that produces the prenylated FMN (pFMN) used as a cofactor by UbiD [25]. Recent studies have provided structural insights into the mechanism of both enzymes, detailing unusual chemistry in each case [26–28]. Crystal structures of UbiD from *E. coli* in complex or not with pFMN have been solved, showing the quarternary structure as homohexamers [27]. The 3D-structure of UbiX from *Pseudomonas aeruginosa* is organized as a homododecamer [25]. Interestingly, Blue Native-PAGE of *E. coli*'s soluble extracts showed a co-migration of UbiD and UbiX at ~700 kDa, compatible with a UbiD$_6$-UbiX$_{12}$ complex (theoretical mass of 582 kDa) suggested by the individual 3D-multimeric structures [29].

Both *O*-methylation reactions in the biosynthesis of UQ$_8$ are catalyzed by the *S*-adenosyl-L-methionine (SAM)-dependent UbiG protein (Figure 1). Crystal structures of UbiG in complex with *S*-adenosyl-L-homocysteine have been determined, with the proteins organized as monomers [30,31]. Interestingly, the conserved region from amino acid 165 to 187 was identified in UbiG as essential for *in vivo* UQ production and for *in vitro* interaction with liposomes. The authors hypothesized that, upon interaction with membrane lipids, this region may promote the entrance of SAM into the protein [30,31]. However, whether or not the membrane association of UbiG contributes to its catalytic activity has not yet been investigated. Moreover, UbiG purified from *E. coli* extracts exhibits *in vitro* methyltransferase activity [32] and a large part of UbiG is detected in the soluble fraction [29]. Thus, the relevance of the lipid binding region of UbiG remains unclear, especially when considering that UbiG is part of the soluble Ubi complex [29] (see 3.3). The *C*-methylation reaction of the pathway is catalyzed by UbiE, a SAM-dependent methyltransferase that is involved in the biosynthesis of UQ and MK [33]. UbiE, for which no structural information is yet available, converts DDMQ$_8$ to DMQ$_8$ (2-octaprenyl-6-methoxy-1,4-benzoquinone to 2-octaprenyl-3-methyl-6-methoxy-1,4-benzoquinone, Figure 1) and demethyl-menaquinone to menaquinone [33].

Finally, three related class A flavoprotein monooxygenases (FMOs) – UbiH, UbiI and UbiF - catalyze hydroxylation reactions on the aromatic ring at carbon atoms C-1, C-5, and C-6, respectively [34,35] (Figure 1). These FMOs use dioxygen as a source of hydroxyl [36] and use the flavin adenine dinucleotide (FAD) to activate O$_2$. UbiI and UbiH seem specific of the position that they modify, whereas UbiF has a broader regio-selectivity since it has a limited ability to hydroxylate C-5 in addition to C-6 [35]. The 3D-structure of a truncated form of UbiI revealed an association as a tetramer, with each monomer containing a typical FAD-binding domain with a Rossman-like β/α/β-fold [35]. It is important to note that *in vitro* assays have still not been developed for most Ubi enzymes, owing in part to the difficulty to obtain isolated purified proteins and to manipulate highly hydrophobic substrates.

### 3.2 Accessory proteins in UQ biosynthesis

Besides the enzymes discussed above, accessory proteins are also involved in UQ biosynthesis. UbiB is an important accessory factor given the nearly complete absence of UQ in *E. coli* mutants lacking a functional *ubiB* gene [37]. UbiB was originally assigned to the C5-hydroxylation step [37], which is now known to depend on UbiI [35]. The UbiB family, composed of bacterial UbiB proteins and of the eukaryotic homologs Coq8-ACDK3/4, belongs to the atypical protein kinase-like family [38]. Biochemical studies of Coq8 and ADCK3 showed that these proteins interact with UQ intermediates and possess ATPase activity but lack kinase activity *in trans* [38,39]. Furthermore, the ATPase activity is stimulated by the interaction with membranes containing cardiolipin and by compounds that resemble UQ intermediates [40]. Overall, UbiB family members were hypothesized to couple the hydrolysis of ATP to the extraction of UQ precursors out of the membrane in order to make them available for UQ biosynthetic enzymes [40], but this hypothetical role remains to be confirmed.

Two other accessory factors, UbiJ and UbiK (formerly YigP and YqiC), were identified recently [24,41]. Cells lacking *ubiJ* show a complete absence of UQ, while *ubiK* mutants retain ~ 20% UQ compared to wild-type. The UQ deficiency is apparent only when the cells are grown in oxic conditions, suggesting that UbiJ and UbiK do not play important functions for UQ biosynthesis under anoxic conditions [24,41,42]. Purified UbiJ and UbiK interact and form an elongated UbiJ$_1$:UbiK$_2$ complex [24] (Figure 1). UbiJ is able to bind UQ biosynthetic intermediates via its SCP2 domain (Sterol Carrier Protein 2), which crystal structure was solved recently [29]. The current hypothesis is that UbiJ and UbiK assist several steps of UQ biosynthesis by presenting UQ intermediates to Ubi enzymes inside the Ubi complex (see 3.3) [29]. In addition to producing a protein, the *ubiJ* locus was proposed to encode a small non-coding RNA (sRNA) termed EsrE [43,44]. EsrE is composed of 252 nucleotides and resides in the 3' half of the *ubiJ* gene [44]. Our group showed that the C-terminal part of the UbiJ protein is sufficient to maintain a minimal level of UQ biosynthesis and we provided evidence to rule out the implication of a sRNA [41]. In contrast, another group reported that both the UbiJ protein and the sRNA EsrE are involved in UQ biosynthesis [44,45]. While some controversy remains, data from both groups agree that the main contribution of the *ubiJ* locus to UQ biosynthesis is carried out by the UbiJ protein rather than by the sRNA EsrE [41,45].

3.3 <u>Supramolecular organization of the *E. coli* UQ biosynthetic pathway</u>

Since UbiA adds the polyprenyl tail onto 4-HB early on (Figure 1), most biosynthetic intermediates of the UQ pathway contain the octaprenyl tail and are therefore highly hydrophobic. Surprisingly, Ubi proteins acting downstream of UbiA are mostly soluble and the last six reactions of the pathway take place in soluble extracts, and not in the membrane fraction as the hydrophobicity of the biosynthetic intermediates would predict [29]. In fact, we showed that a ~1 MDa Ubi complex composed of seven proteins (UbiE-K) exists in the soluble fraction of *E. coli* extracts [29]. This complex contains the five enzymes (UbiE-I) that catalyze the reactions downstream of OHB (Figure 1) and the accessory factors UbiJ and UbiK, the former being essential to the stability of the Ubi complex [29]. We also demonstrated that the biosynthetic intermediates OPP and DMQ$_8$ are bound in the Ubi complex and we proposed that the N-terminal SCP2 domain of UbiJ mediates the interaction [29]. Altogether, the current model is that the Ubi complex forms a soluble metabolon that synthesizes UQ from OHB (Figure 1). The trafficking of these two hydrophobic molecules between the membrane and the Ubi complex might involve the UbiB protein with its ATPase activity and its predicted C-terminal transmembrane domain [29,40].

Interestingly, a similar organization of the UQ pathway has also been described in eukaryotes with complex Q (also termed the 'CoQ-synthome'). Complex Q groups the enzymes of the late steps [10,46], but it is associated to the membrane, contrary to the Ubi complex which is soluble. Outstanding questions remain regarding the supramolecular assembly of the UQ pathway, notably the conservation of the Ubi complex in other bacterial species, the exact composition and stoichiometry of the complexes, their 3D structures, their potential dynamic nature and their cellular localization. A recent study started to address the two latter points in yeast [47].

### 3.4 Discovery of a conserved O$_2$-independent pathway

Based on the observation that *E. coli* was able to synthesize UQ under anoxic conditions, the existence of a UQ biosynthesis pathway independent from O$_2$ had long been hypothesized [48]. This pathway remained uncharacterized until 2019, when we identified three genes, *ubiT*, *ubiU* and *ubiV* which are required for UQ biosynthesis under anoxic conditions but are dispensable under oxic conditions [42]. The only reactions that differ between the O$_2$-dependent and O$_2$-independent pathways are the three hydroxylation steps catalyzed by the O$_2$-consuming flavin hydroxylases UbiI, UbiH and UbiF [42]. UbiU and UbiV, which belong to the U32 peptidase family, form a heterodimer that is required for the hydroxylation of DMQ$_8$ *in vivo.* This result is in the line with the demonstration that UbiU from *P. aeruginosa* co-purifies with UQ$_8$ and DMQ$_8$ [49] and we postulated that UbiU and UbiV may participate to the two other hydroxylation steps of the anaerobic UQ pathway [42,49]. A role for UbiU and UbiV in O$_2$-independent hydroxylation reactions is supported by recent studies showing that two other U32 peptidase family members - RlhA and TrhP – are required for the hydroxylation of C2501 on 23S rRNA [50] and of U34 on some tRNAs [51,52], respectively. The source of oxygen used in the hydroxylation reactions involving U32 peptidase family members is unknown at this stage but prephenate, a metabolite of the shikimate pathway, is a candidate since it is required for the function of RlhA and TrhP [50,52]. The presence of an iron-sulfur cluster might be another feature common to U32 proteins. Indeed, we showed that UbiU and UbiV each carry a 4Fe-4S cluster ligated by a motif of conserved cysteine residues, which is found in most U32 peptidase family members [42]. Interestingly, the function of RlhA and TrhP depends on these Cys residues and on the genes of the *isc* operon that catalyze the biogenesis of Fe-S clusters [50,52]. Some additional players may also be involved in the function of UbiU and UbiV, like the low potential ferredoxin YhfL, which is required for the hydroxylation of tRNAs by TrhP [51]. Overall, the U32 peptidase family emerges as a new class of $O_2$-independent hydroxylases and additional work is required to elucidate the mechanism of these enzymes and the precise function of UbiU and UbiV in UQ biosynthesis.

The role of UbiT in the $O_2$-independent UQ biosynthetic pathway is still unclear. Yet, the presence of a SCP2 domain in the sequence of UbiT and the demonstration that UbiT binds the lipid phosphatidic acid [53] suggests that UbiT's function is linked to lipids. Moreover, we recently showed that UbiT from *P. aeruginosa* binds $UQ_8$ by recognizing its isoprenoid tail [49], suggesting that UbiT may perform a role similar to UbiJ in presenting the hydrophobic intermediates of the UQ pathway to Ubi enzymes. Interestingly, UbiJ is important for UQ biosynthesis only in oxic conditions, whereas the role of UbiT is limited to anoxic conditions [41,42]. The possibility that UbiJ and UbiT may functionally replace each other depending on environmental conditions is an appealing hypothesis, given that both SCP2 proteins need to assist different sets of UQ biosynthetic enzymes, the $O_2$-dependent hydroxylases (UbiI, UbiH, UbiF) in one case, and the $O_2$-independent hydroxylases (likely UbiU and UbiV) in the other case.

The *ubiT*, *ubiU* and *ubiV* genes are widespread in proteobacterial genomes that possess the $O_2$-dependent UQ pathway, suggesting that numerous bacteria have the previously unrecognized capacity to synthesize UQ over the entire $O_2$ range [42]. The low potential menaquinone (MK) is typically involved in transferring electrons in anaerobic respiratory chains, thus the physiological function(s) of UQ synthesized in anoxic conditions remains to be clarified in proteobacteria possessing both UQ and MK pathways. Several gram-negative bacteria, such as *P. aeruginosa*, contain UQ as sole quinone [2]. We found that the *ubiT*, *ubiU* and *ubiV* genes are essential for UQ production by *P. aeruginosa* in anoxic conditions and that these genes are required for denitrification [49], a metabolism on which *P. aeruginosa* heavily relies to develop in the lungs of cystic fibrosis patients. Overall, the discovery of a widespread UQ pathway independent of $O_2$ certainly changes our perspective of the relative contribution of various quinones to bacterial metabolisms in hypoxic and anoxic conditions. Interestingly, substantial amounts of UQ were reported lately in the anoxic zone of the water column of the Black Sea [5], suggesting that an $O_2$-independent pathway could have been at work in this ecosystem. It remains to be investigated whether or not bacteria containing *ubiT*, *ubiU* and *ubiV* genes are found in this ecological niche. By extension, assessing the contribution of the $O_2$-independent UQ pathway to anaerobiosis constitutes an exciting new research avenue.

**4) Variations in UQ biosynthesis pathways across bacteria**

Our current view of the biosynthesis of UQ in bacteria is mostly based on the *E. coli* pathway [15]. Even though numerous discoveries on *E. coli* are applicable to other bacterial species, recent studies using other bacterial models revealed an unsuspected diversity in the composition of the UQ biosynthesis pathway across bacteria.

4.1 Synthesis of the aromatic ring precursor

So far, only 4-HB has been described as an aromatic ring precursor for UQ in bacteria. In contrast, eukaryotes are able to use additional molecules like para-aminobenzoic acid (pABA) [54,55]. Note that pABA was shown to be processed through several steps of the UQ pathway in *E. coli* [55]. However, the amino-substituted intermediates were not converted into UQ [55], thus pABA is not considered a precursor for UQ in *E. coli*. The first gene identified to synthesize 4-HB for bacterial UQ biosynthesis was *ubiC*, which encodes a chorismate pyruvate-lyase [20]. The *xanB2* gene of *Xanthomonas campestris* was later shown to encode a chorismatase that produces 4-HB for UQ biosynthesis and 3-hydroxybenzoic acid for the biosynthesis of pigments from the xanthomonadin family [56]. Even though UbiC and XanB2 use the same substrate – chorismate, the end product of the shikimate pathway – they do not share sequence or structural identities and belong to different protein families, chorismate pyruvate-lyase and chorismatase, respectively [56]. *xanB2* is present in several proteobacterial genera that do not contain *ubiC* [56], supporting a strong anti-occurrence of the two genes, although this has not been analyzed in details. It is currently unclear if all UQ producing bacteria contain UbiC or XanB2 or if additional unidentified 4-HB generating systems might also be involved in some species. Interestingly, a new subfamily of chorismatase (type IV) was shown to produce only 4-HB (and not a mixture of 3-HB and 4-HB as the type III chorismatase XanB2) [57] and may therefore represent a new candidate to produce 4-HB for UQ biosynthesis.

4.2 Hydroxylases

Three hydroxylation reactions on contiguous positions of the aromatic ring are required during the biosynthesis of UQ. The enzymes (UbiI, UbiH and UbiF) involved in the $O_2$-dependent *E. coli* pathway each hydroxylate one position and belong to the same family of flavin monooxygenases. An unrelated di-iron monooxygenase Coq7 is implicated in the C-6 hydroxylation instead of UbiF in some bacterial species [58–60]. In 2016, a search for these four monooxygenases over representative proteobacterial genomes led to the identification of two new flavin monooxygenases, UbiL and UbiM [61]. This study revealed an astonishing diversity of combinations of monooxygenases used by bacteria to synthesize UQ (19 combinations in 67 species) [61]. Interestingly, some genomes contained less than three UQ monooxygenases [61]. We demonstrated that the UbiL protein from *Rhodospirillum rubrum* hydroxylates two positions (C-1 and C-5) and that the UbiM protein from *Neisseria meningitidis* hydroxylates three positions, rationalizing the presence of respectively two and one UQ monooxygenase genes in these species. Some genes are restricted to specific classes (*ubiL* to α- and *ubiF* to γ-proteobacteria), while the distribution of *ubiM* across α, β, γ-proteobacteria is likely the result of horizontal gene transfer [61]. Intriguingly, some species such as *Xanthomonas campestris* or *Alteromonas macleodii* contain four UQ monooxygenases [61]. The reason as to why bacteria evolved such a diversity of $O_2$-dependent UQ monooxygenases is still unknown. Of note, the putative hydroxylases of the $O_2$-independent pathway show probably less diversity since a very high co-occurrence of UbiU and UbiV was observed [42].

### 4.3 Incomplete UQ biosynthesis pathways

The decarboxylation step of the pathway seems also variable. Indeed, the only enzyme implicated so far is the UbiD decarboxylase assisted by the prenyl-transferase UbiX [62]. However, several authors recently noticed the absence of *ubiX-ubiD* genes from genomes containing most of the other *ubi* genes, suggesting that another enzymatic system could be involved in the decarboxylation reaction [60,63,64]. A candidate gene *ubiZ* was proposed based on its co-localization with *ubiE* and *ubiB* in the genomes of *Acinetobacter spp.* and *Psychrobacter sp.* PRwf-1 [64]. The function of UbiZ remains to be investigated, but the fact that the *ubiZ* gene is not conserved in all the genomes lacking *ubiD* and *ubiX* suggests the existence of yet another decarboxylation system in UQ biosynthesis (Table 1).

Another intriguing possibility is that incomplete quinone biosynthesis pathways might nevertheless be functional. Indeed, organisms with incomplete pathways might be able to scavenge particular metabolites from their environment rather than to synthesize them intracellularly. As such, genetic gaps in *Wolbachia* for the biosynthesis of 4-HB and of isopentenyl-pyrophosphate (one of the building blocks of the polyprenyl tail of UQ), led the authors to propose that these compounds might be acquired exogenously in order to support UQ biosynthesis [65]. Remarkably, *Streptococcus agalactiae* synthesizes its demethylmenaquinone thanks to a partial MK biosynthesis pathway and thanks to the uptake of the late intermediate 1,4-dihydroxy-2-naphthoic acid (DHNA) from the extracellular environment [66]. Several *Lactobacillus* species also contain a partial MK pathway [64], suggesting that these bacteria might also rely on the import of exogenous intermediates to synthesize MK. Exchanges of metabolites between species are common in bacterial communities as in the gut of vertebrates, and small soluble components like DHNA are likely exchanged more easily than the large hydrophobic intermediates of the UQ pathway. Therefore, this strategy of complementing a partial pathway by importing extracellular intermediates is certainly more applicable to quinone pathways with a prenylation reaction occurring at a late stage (like the MK pathway in which most intermediates are small and hydrophilic [7]) rather than to UQ and PQ pathways with early prenylation steps, and consequently large and hydrophobic intermediates.

Overall, the large diversity of combination of enzymes used to synthesize UQ in various environmental conditions (Table 1) leads us to refer to UQ biosynthesis pathway*s* and not anymore to a single pathway, as already proposed by Degli Esposti [63]. We envision that even more UQ pathways will be revealed by systemic bioinformatic approaches aimed at studying the variations of UQ biosynthesis in the ever expanding diversity of bacterial genomes available. Let's mention here that the task faces several difficulties, one of which is that only some *ubi* genes tend to group into operonic structures whereas others are dispersed around the chromosome [15].

**Table 1:** Protein composition of the bacterial UQ biosynthetic pathways. *italics*: proteins involved only in the $O_2$-independent pathway; underlined: proteins involved only in the $O_2$-dependent pathway, ?: suspected existence of unidentified alternative proteins

| | *E. coli* pathways | | Alternative proteins in other bacteria |
|---|---|---|---|
| **Step or function** | $O_2$-dependent | $O_2$-independent | |
| Synthesis of polyprenyl-pyrophosphate | IspA, IspB | IspA, IspB | |
| Synthesis of 4-HB | UbiC (chorismate lyase) | UbiC | XanB2 (chorismatase) |
| Polyprenyl transferase | UbiA | UbiA | |
| Decarboxylation | UbiD (decarboxylase) UbiX (flavin prenyltransferase) | UbiD UbiX | UbiZ, ? |
| Methylation | UbiG, UbiE | UbiG, UbiE | |
| Hydroxylation | UbiH, UbiF, UbiI (flavin cofactor) | *UbiU, UbiV* (Fe-S cofactor) | UbiL, UbiM (flavin cofactor) Coq7 (di-iron cofactor) |
| SCP2 protein | UbiJ | *UbiT* | |
| ATPase | UbiB | UbiB | |
| Accessory factor | UbiK | | |

## 5) An evolutionary perspective on (ubi)quinone biosynthetic pathways

The rise of $O_2$ concentrations on Earth caused a shift from globally reducing to oxidizing conditions around 2.4 billion years ago [67]. This transition had far-reaching consequences, notably for quinones. Indeed, the low potential MK, which was present at the time of the great oxidation event, is readily oxidized by $O_2$ [68]. Thus, it was proposed that microorganisms had to evolve higher potential quinones, like UQ and PQ, to sustain electron transport in bioenergetic chains operating under oxidizing conditions [68]. This scenario is in line with the presence of $O_2$-requiring steps, respectively three and one, in the biosynthetic pathways for UQ and PQ (Figure 2). However, our recent discovery of an $O_2$-independent pathway for UQ production, widespread across proteobacterial lineages, suggests that UQ biosynthesis might have emerged in a less favorable $O_2$ context than previously thought [42]. One way to tackle the question of the relative origins of the quinone pathways is to study the evolution of the involved enzymes provided homologs are shared between pathways.

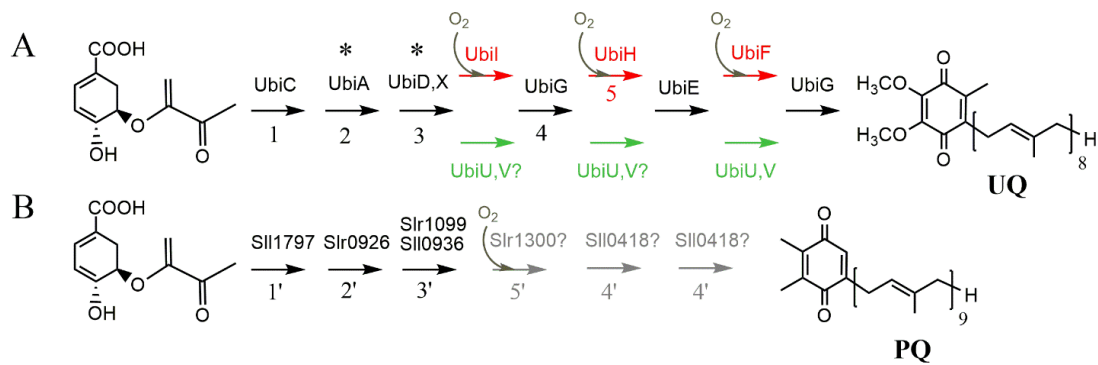

**Figure 2**: Homology between bacterial UQ- and cyanobacterial PQ- pathways. A) Biosynthetic pathway of UQ in *Escherichia coli* with enzymes specific of the $O_2$-dependent and $O_2$-independent pathways in red and green respectively. B) Biosynthetic pathway of PQ in the cyanobacterium *Synechocystis sp.* PCC 6803. Reactions 1-3 in the UQ pathway and 1'-3' in the PQ pathway are catalyzed by homologous enzymes. Proposed candidates for the PQ pathway (Slr1300? and Sll0418?) are homologous to UbiH and UbiG (see text). Enzymes with homologs in MK pathways are designated with (*).

### 5.1 Evolution of the UQ and PQ pathways

It should be possible to address the relative appearance of the UQ and the PQ pathways since they share several homologs. Here, we consider only the cyanobacterial PQ pathway which consist of six reactions (Figure 2), as opposed to the pathway found in plant which is entirely different [69]. In the cyanobacterium *Synechocystis* sp., the first three steps of PQ biosynthesis involve homologs to UbiC, UbiA, and UbiD – UbiX of the UQ pathway: respectively, the chorismate lyase Sll1797, the 4-HB prenyltransferase Slr0926, and the decarboxylase - flavin prenyltransferase Sll0936 - Slr1099 [70,71] (Figure 2). The following hydroxylation and methylation steps are still to be experimentally validated, but candidates have been proposed (Slr1300 and Sll0418) based on their homology to UbiH and UbiG enzymes of the UQ pathway [72]. Degli Esposti conducted a phylogenetic analysis of the UbiA, -C, -D, -H homologs and proposed that the UQ pathway derived from the PQ pathway and appeared twice independently in Alphaproteobacteria and in Zetaproteobacteria [63]. Yet the trees built in this study are missing outgroups to root the phylogenies and as such do not definitively address the question of the relative origins of the UQ and PQ pathways [63].

### 5.2 Relationships between the UQ and the MK pathways

Two pathways are known for the biosynthesis of MK [7]: a fully characterized, long-known "classical MK pathway" and a still incomplete, more recently identified "futalosine pathway" [73]. The classical MK pathway has only two steps in common with the UQ pathway: the prenylation step catalyzed by the prenyltransferase MenA (homologous to UbiA) and the methylation of the aromatic ring catalyzed by the literally shared enzyme MenG/UbiE. The characterized *mqnA-E* genes are specific to the futalosine pathway [74], but the still putative MqnP, MqnL, MqnM, and UbiE/MenG have homologs in the UQ pathway (UbiA, UbiD and UbiX, respectively) [64,73]. These later *mqnP, -L, -M* genes were found to strictly co-occur with *mqnA-E* in many bacterial genomes, which reinforces their potential to participate in the futalosine pathway [64]. In 2014, Zhi and colleagues observed that the futalosine pathway was found in more phyla of Bacteria and Archaea than the classical MK pathway [9]. Furthermore, phylogenies of MenB, -C, -F suggested that the classical MK pathway was acquired in Archaea, specifically in Halobacteriaceae, as a result of lateral gene transfers from bacteria. In contrast, phylogenies for the MqnA, -D, and -C enzymes (specific to the futalosine pathway) globally retrieved the delineation of major bacterial and archaeal lineages, suggesting a vertical inheritance of the futalosine pathway and an early emergence predating that of the classical pathway [9].

In 2016, Ravcheev and Thiele built phylogenies for genes of the two MK, and the UQ pathways [64]. Their trees showed that homologs from the different pathways separated well (including those of the candidate MqnP, -L and –M, homologs of UQ enzymes). Interestingly, the only enzyme supposedly shared by the three pathways, the prenyltransferase UbiA/MenA/MqnP family, had a phylogeny displaying a dichotomy between the classical MK pathway on one side, and the futalosine and UQ pathways on another side [64].

However, in the tree of the methyltransferase family (UbiE/MenG), candidate enzymes of the futalosine pathway positioned within those of the classical MK pathway, and apart from those of the UQ pathway. The authors therefore suggested that the likely younger pathway of UQ evolved from parts of the two pre-existing MK pathways, with some enzymes being more closely related to the futalosine pathway, and others to the classical pathway.

Future studies in the context of recent discoveries, including that of new pathways (e.g. the $O_2$-independent UQ pathway) or new taxonomic groups of Archaea and Bacteria [75], are very likely to further enlighten the origins of quinones. Understanding the evolutionary relationships of the quinone pathways is indeed important as it bears strong implications for understanding the evolution of bioenergetics and adaptation to extant oxidizing environments.

## 6) Conclusion and Perspectives

Recent results have significantly expanded our view of the biosynthesis of UQ in bacteria. Several functional homologs have now been identified at various steps (Table 1) and a pathway independent from $O_2$ has been characterized [42]. The first proof of a supramolecular structuration of the *E. coli* $O_2$-dependent UQ pathway was recently provided with the characterization of the Ubi complex [29]. Whether such multiprotein complexes exist or not in other bacterial species and how they accommodate the variability of the constituting proteins (notably the hydroxylases) remains to be investigated. Understanding the regulation of the various UQ pathways and establishing their cellular localization will also be of interest. Indeed, we may expect the UQ biosynthesis apparatus to localize close to active bioenergetic enzymes, and some of them adopt a specific localization, as recently observed for the fumarate dehydrogenase and nitrate reductase in respiring *E. coli* [76]. Whether the UQ pathways indeed originated from the MK pathways [64] and how they evolved in the past 2 billion years is also a challenging and interesting question.

To obtain a satisfactory understanding of the composition, regulation and evolution of the UQ pathways across bacteria, it will certainly be fruitful to combine biochemical and bioinformatic approaches in order to extract information from the multiple genomes now available in public databases. Besides increasing our basic knowledge of UQ pathways, such studies will also benefit bioengineering projects aimed at increasing the production of UQ [77] or that of related natural products like antroquinonol, a molecule currently in clinical trials for non-small-cell lung cancer [78]. In addition, a better understanding of the UQ pathways may refine possible strategies to target them in order to develop novel antibiotics, and may also provide valuable information to help pinpoint the nature of the bacterial ancestor of mitochondria [79].

**Acknowledgments**: This work was supported by the Agence Nationale de la Recherche (ANR), projects (An)aeroUbi ANR-15-CE11-0001-02, O2-taboo ANR-19-CE44-0014, by the Grenoble Alpes Data Institute funded under the "Investissements d'avenir" program (ANR-15-IDEX-02), by the Centre National de la Recherche Scientifique (CNRS) and by the University Grenoble Alpes (UGA).